# Influence of impurity on isotope coefficient of superconductors

P. Udomsamuthirun [1]

[1] Department of Physics, Faculty of Science, Srinakharinwirot University, Bangkok 10110, Thailand. E-mail: udomsamut55@yahoo.com

**Abstract**

We study influence of anisotropy of the impurity potential on isotope coefficient of impure superconductor in Born limit. The isotope coefficient of nonmagnetic and magnetic impurities including an angle-resoled Fermi surface density of state, anisotropy order parameter and impure potential that we make the assumption that the form of them are in the harmonic form are investigated. Our numerical calculation of isotope coefficient can fit well with experimental data of high-$T_c$ superconductors.

## 1. Introduction

One of the important parameters in the experiment and theory of superconductor is the isotope coefficient $\alpha$ defined by $T_c \alpha M^{-\alpha}$ where $T_c$ is the superconducting transition temperature and $M$ is the isotope mass of the element. In the BCS theory it has been predicted that $\alpha = 0.5$ suggesting phonon-mediated pairing in superconductors that verified experimentally for conventional superconductors. In high temperature superconductors the isotope effect shows complex behavior : some measurements[1,2] have shown near zero isotope effect coefficient and some measurements[3,4] have shown the values greater than 0.5 .To explain the unusual isotope effect in high temperature superconductors, many models have been proposed such as the van Hove singularity[5-7], anharmonic phonon[8,9], pairing-breaking effect[10] and pseudogap[11,12], nonadiabaticity[13] and impurity effect[14,15].

The results of impurities that introduced into the superconductors are found in a change of the superconducting transition temperature and the isotope coefficient. They modify the quasi-particle spectrum, interaction parameters and induce pair breaking in superconducting state. The non-magnetic impurities have little effect on transition temperature in s-wave superconductors [16] but they exhibit a strong pair breaking effect in the high temperature superconductors [17] . Shi and Li[18] study the influence of nonmagnetic impurities on transition temperature of the layered superconductors by using an superconducting-normal layer model with d-wave pairing in the superconducting layer. Mierzynska and Wysokinski[15] study the effect of in-plane and out-of-plane impurities on isotope effect coefficient in layered superconductors. Haran and Nagi [14,19,20,21] proposed a theory of nonmagnetic and magnetic impurities in an anisotropic superconductors including the effect of anisotropic (momentum-dependent) impurities scattering. Openov , Semenihin and Kishore [22] study the effect of impurities, both magnetic and nonmagnetic, on the isotope effect coefficient in high temperature superconductors in the framework of Abrikosov-Gorkov approach. But Openov et al.[22] do not include the effect of momentum-dependent of scattering potential. Although Haran and Nagi consider the effect of anisotropic (momentum-dependent) impurities scattering, they do not consider the form of anisotropic function and apply to experimental data.

The purpose of this work is to apply the model of nonmagnetic and magnetic impurity of Haran and Nagi to investigate the influence of an angle-resoled Fermi surface density of state, anisotropic order parameter and anisotropic impure potential on isotope effect coefficient of superconductors. To apply this model, we make the assumption that an angle-resoled Fermi surface density of state and all of potentials of this model are in the harmonic form. Finally we show the numerical calculation of isotope coefficient that fit well with experimental data of high-$T_c$ superconductors.

**2. The isotope coefficient**

Haran and Nagi [14,19,20,21] consider the problem of nonmagnetic and magnetic impurities in an anisotropic superconductor for the case of anisotropic (momentum-dependent) impurity scattering in weak-coupling approximation. They consider an anisotropic superconductor with randomly distributing impurities, treating the electron-impurity scattering within second Born approximation, and neglecting the impurity-impurity interaction. The normal and anomalous temperature Green's function averaged over the impurity position read

$$G(\omega,k) = -\frac{i\tilde{\omega}+\xi_k}{\tilde{\omega}^2+\xi_k^2+\left|\tilde{\Delta}(k)\right|^2} \quad (1)$$

$$F(\omega,k) = \frac{\tilde{\Delta}(k)}{\tilde{\omega}^2+\xi_k^2+\left|\tilde{\Delta}(k)\right|^2} \quad (2)$$

where the renormalized Matsubara frequency $\tilde{\omega}(k)$ and the renormalized order parameter $\tilde{\Delta}(k)$ are given by

$$\tilde{\omega}(k) = \omega + in_i \int \left|u(k,k')\right|^2 G(\omega,k')\frac{d^3k'}{(2\pi)^3} \quad (3)$$

$$\tilde{\Delta}(k) = \Delta(k) + in_i \int \left|u(k,k')\right|^2 F(\omega,k')\frac{d^3k'}{(2\pi)^3} \quad (4)$$

Here $\omega = \pi T(2n+1)$, T is temperature, $n$ is integer number. $\xi_k$ is the quasiparticle energy, $n_i$ is impurity concentration, $u(k,k')$ is a momentum-dependent impurity potential, and $\Delta(k)$ is the orbital part of the singlet superconducting order parameter defined as $\Delta(k) = \Delta\, e(k)$ where $e(k)$ is a real basis function and $<e^2>=1$, where $<..>$ denotes the average value over the Fermi surface that

$$\int \frac{d^3k}{(2\pi)^3} \to N_0 \int_{FS} dS_k n(k) \int d\xi_k \quad \text{and} \quad <..> = \int_{FS} dS_k n(k)(..). \text{ Here } n(k) \text{ is the angle-}$$

resolved Fermi surface density of state. $N_0$ is the overall density of state at Fermi surface. The momentum-dependent impurity potential is

$$u(k,k') = v(k,k') + J(k,k')\vec{S} \cdot \vec{\sigma} \tag{5}$$

where $\vec{S}$ is a classical spin of impurity and $\vec{\sigma}$ is the electron spin density and assumed a separable form of scattering probabilities.

$$v^2(k,k') = v_0^2 + v_1^2 f(k)f(k') \tag{6}$$

$$J^2(k,k') = J_0^2 + J_1^2 g(k)g(k') \tag{7}$$

where $v_0(v_1), J_0(J_1)$ are isotropic (anisotropic) scattering amplitudes for non-magnetic and magnetic potential. $f(k), g(k)$ are the momentum-dependent anisotropy function in the nonmagnetic and magnetic scattering channel and let $g(k) = \pm f(k)$ The averaged over the Fermi surface of $f(k)$ and $g(k)$ vanish,

$<f(k)> = <g(k)> = 0$, and are normalized as $<f^2(k)> = 1$. And $\Gamma_0 = \pi n_i N_0 v_0^2$, $\Gamma_1 = \pi n_i N_0 v_1^2$, $G_0 = \pi n_i N_0 J_0^2 S(S+1)$, $G_1 = \pi n_i N_0 J_1^2 S(S+1)$ are scattering rate of isotropic non-magnetic, anisotropic non-magnetic, isotropic magnetic, and anisotropic magnetic channel, respectively.

Substitution Eqs.(5-7) into Eqs.(3,4), we can get the transition temperature as [14]

$$\ln(\frac{T_C}{T_{C0}}) = (1 - <e>^2 - <ef>^2)[\Psi\left(\frac{1}{2}\right) - \Psi\left(\frac{1}{2} + (\frac{\Gamma_0 + G_0}{2\pi T_C})\right)] + <e>^2 [\Psi\left(\frac{1}{2}\right) - \Psi\left(\frac{1}{2} + (\frac{2G_0}{2\pi T_C})\right)]$$

$$+ <ef>^2 [\Psi\left(\frac{1}{2}\right) - \Psi\left(\frac{1}{2} + (\frac{\Gamma_0 + G_0 + G_1 - \Gamma_1}{2\pi T_C})\right)] \tag{8}$$

Here $T_C$ ($T_{C0}$) is the critical temperature of impure(clean) superconductors.

After some calculation, the isotope effect coefficient of impure superconductors in the Born approximation is found as[14]

$$\frac{\alpha_0}{\alpha} = 1 - (1 - <e>^2 - <ef>^2)\left(\frac{\Gamma_0 + G_0}{2\pi T_C}\right)\Psi'\left(\frac{1}{2} + (\frac{\Gamma_0 + G_0}{2\pi T_C})\right) - <e>^2 \left(\frac{2G_0}{2\pi T_C}\right)\Psi'\left(\frac{1}{2} + (\frac{2G_0}{2\pi T_C})\right)$$

$$- <ef>^2 \left(\frac{\Gamma_0 + G_0 + G_1 - \Gamma_1}{2\pi T_C}\right)\Psi'\left(\frac{1}{2} + (\frac{\Gamma_0 + G_0 + G_1 - \Gamma_1}{2\pi T_C})\right) \tag{9}$$

Here $\alpha$ ($\alpha_0$) is the isotope coefficient of impure(clean) superconductors.

Eq.(8) and Eq.(9) are the general equation of $T_C$ and isotope coefficient $\alpha$ that contain both nonmagnetic and magnetic impurities, anisotropy of order parameter, and anisotropy of scattering potential of impurities. They can be reduced to nonmagnetic impurity case [19] by set $G_0 = G_1 = 0$. In the work of Opennov, Semenihim and Kishore [22], they also study the critical temperature and isotope coefficient of anisotropy superconductor containing both nonmagnetic and magnetic impurities. The coefficient $\chi = 1 - \frac{<\Delta(p)>_{FS}^2}{<\Delta(p)^2>_{FS}}$ of their model can be reduced to $\chi = 1 - <e>^2$. The total electron relaxation time due to potential scattering by both magnetic and nonmagnetic impurities $\tau = \frac{2}{\Gamma_0 + G_0}$ and the electron relaxation time due to exchange scattering by magnetic impurities $\tau_m^{ex} = \frac{2}{G_0}$. The second term on the right-side of Eq.(8) do not appear in Opennov, Semenihim and Kishore's $T_C$ equation because they do not consider the momentum-dependent of impurity scattering potential.

### 3. Calculation and discussion

To show the numerical calculation of this model, we must know the form of function $e(k)$, $f(k)$ and $n(k)$. Firstly, we consider the form of order parameters. The order parameters of s-wave superconductors are corresponded to $\Delta(k) = \Delta_0$ that give $e(k) = 1$. And for the high temperature superconductors, they are most likely layered $d_{x^2-y^2}$-wave pairing symmetry. In $d_{x^2-y^2}$-wave superconductor the order parameters are corresponded to $\Delta(k) = \Delta_0 \cos 2\phi$, that give $e(k) \equiv \cos 2\phi$. Secondly, we consider the form of the anisotropic impurity scattering potential. The possible of the anisotropy impurity scattering potential was suggested in a discussion of the irradiation data by Giapintzakis et al.[23]. Haran and Nagi[21] take the function $f(k)$ in the scattering potential proportional to the harmonic functions which in a polar notation $\sin(n\phi)$ and $\cos(n\phi)$, where n is an integer number. They use $f(k) \equiv \sin 2\phi$, $\sin 3\phi$, $\cos 2\phi$, and higher-order harmonics to study the critical temperature of impure superconductor in Born limit. Thirdly, we consider the form of an angle-resoled Fermi surface density of state. The model of an angle-resoled Fermi surface density of state,

$n(\phi) \equiv 1$, $n(\phi) \equiv \cos 2\phi$ and $n(\phi) \equiv \cos 4\phi$, are used by Mierzynska and Wysokinski [15] in the study of the effect of in-plane and out-of-plane impurities on isotope effect coefficient in layered superconductors.

As corresponding to that mention in above paragraph, we make the assumption that the form of order parameters, the form of the anisotropic impurity scattering potential and the form of an angle-resoled Fermi surface density of state ($e(k)$, $f(k)$ and $n(k)$) are in the harmonic form. The constraint conditions are $<e^2>=1$, $<f(k)>=0 < f^2(k)>=1$, and $\int_{FS} dS_k n(k) = 1$. Within these constraints and our assumptions, we can show the values of $<e>$ and $<ef>$ of s-wave, d-wave order parameters in table below.

| Order parameter | $n(\phi)$ | $e(\phi)$ | $f(\phi)$ | $<e>$ | $<ef>$ |
|---|---|---|---|---|---|
| s-wave | $\dfrac{1}{2\pi}$ | 1 | $\sqrt{2}\cos\phi$ | 1 | 0 |
| | | | $\sqrt{2}\cos 2\phi$ | 1 | 0 |
| | | | $\dfrac{4}{\sqrt{5}}\cos^3 2\phi$ | 1 | 0 |
| | $\dfrac{1}{\pi}\cos^2\phi$ | 1 | $\dfrac{2}{\sqrt{3}}\cos\phi$ | 1 | 0 |
| | | | $2\sin\phi$ | 1 | 0 |
| | $\dfrac{1}{\pi}\sin^2\phi$ | 1 | $\dfrac{2}{\sqrt{3}}\sin\phi$ | 1 | 0 |
| | | | $2\cos\phi$ | 1 | 0 |
| $d_{x^2-y^2}$-wave | $\dfrac{1}{2\pi}$ | $\sqrt{2}\cos 2\phi$ | $\sqrt{2}\cos 2\phi$ | 0 | 1 |
| | | | $\dfrac{4}{\sqrt{5}}\cos^3 2\phi$ | 0 | $\dfrac{3}{\sqrt{10}}$ |
| | | | $\dfrac{16}{\sqrt{63}}\cos^5 2\phi$ | 0 | $\dfrac{5}{3}\sqrt{\dfrac{2}{7}}$ |

| $\frac{1}{\pi}\cos^2\phi$ | $\sqrt{2}\cos 2\phi$ | $2\sin\phi$ | $\frac{1}{\sqrt{2}}$ | 0 |
|---|---|---|---|---|
| | | $\frac{2}{\sqrt{3}}\cos\phi$ | $\frac{1}{\sqrt{2}}$ | 0 |
| | | $\sqrt{2}\sin 2\phi$ | $\frac{1}{\sqrt{2}}$ | 0 |
| $\frac{1}{\pi}\cos^2 2\phi$ | $\frac{2}{\sqrt{3}}\cos 2\phi$ | $\frac{2}{\sqrt{3}}\cos 2\phi$ | 0 | 1 |
| | | $\frac{8}{\sqrt{35}}\cos^3 2\phi$ | 0 | $2\sqrt{\frac{5}{21}}$ |
| $\frac{1}{\pi}\sin^2\phi$ | $\sqrt{2}\cos 2\phi$ | $\frac{2}{\sqrt{3}}\sin\phi$ | $\frac{1}{\sqrt{2}}$ | 0 |
| | | $2\cos\phi$ | $\frac{1}{\sqrt{2}}$ | 0 |
| | | $\sqrt{2}\sin 2\phi$ | $\frac{1}{\sqrt{2}}$ | 0 |

Substitution the values of $<e>$ and $<ef>$ into Eq.(8) and Eq(9), we can get the isotope coefficient of impure superconductors. We find that the non-magnetic impurities have no effect on isotope coefficient in s-wave superconductors; $\alpha = 0.5$ for all values of parameters and there is little effect of $<e>$ and $<ef>$ on isotope coefficient. Our numerical calculations of isotope coefficients fitted to experimental data [3,4,24] are shown in Figure 1. In this figure, we show the curve of d-wave ($<e>=0, <ef>=1$) with non-magnetic impurity ($\Gamma_0 \neq 0, \Gamma_1 = G_0 = G_1 = 0$) case and s-wave ($<e>=1, <ef>=0$) and d-wave($<e>=0, <ef>=1$) with magnetic impurity ($G_0 \neq 0, \Gamma_0 = \Gamma_1 = G_1 = 0$) case that shown the extremum values of isotope effect coefficient. Most of experimental data are in the region of isotope coefficient between d-wave with non-magnetic impurity case to s-wave and d-wave with magnetic impurity case. The values of isotope coefficient in between this region may be occurred by the mixing of non-magnetic and magnetic impurities.

## 4. Conclusion

We have applied the model of nonmagnetic and magnetic impurity of Haran and Nagi to investigate the influence of an angle-resoled Fermi surface density of state, anisotropy order parameter and impure potential on isotope effect coefficient of impure superconductors. We make the assumption that the form of order parameters, the form of the anisotropic impurity scattering potential and the form of an angle-resoled Fermi surface density of state ($e(k)$, $f(k)$ and $n(k)$) are in the harmonic form. Our numerical calculation can fit well with the experimental data of high-$T_c$ superconductors. We find that there is an effect of nonmagnetic impurity on isotope coefficient in d-wave superconductors but no effect on s-wave superconductors and an effect of magnetic impurity on isotope coefficient are found both in s-wave and d-wave superconductors.

## Acknowledgement

The author would like to thank Professor Dr.Suthat Yoksan for the useful discussion and also thank Thai Research Fund and Office of Higher Education Commission for the financial support.

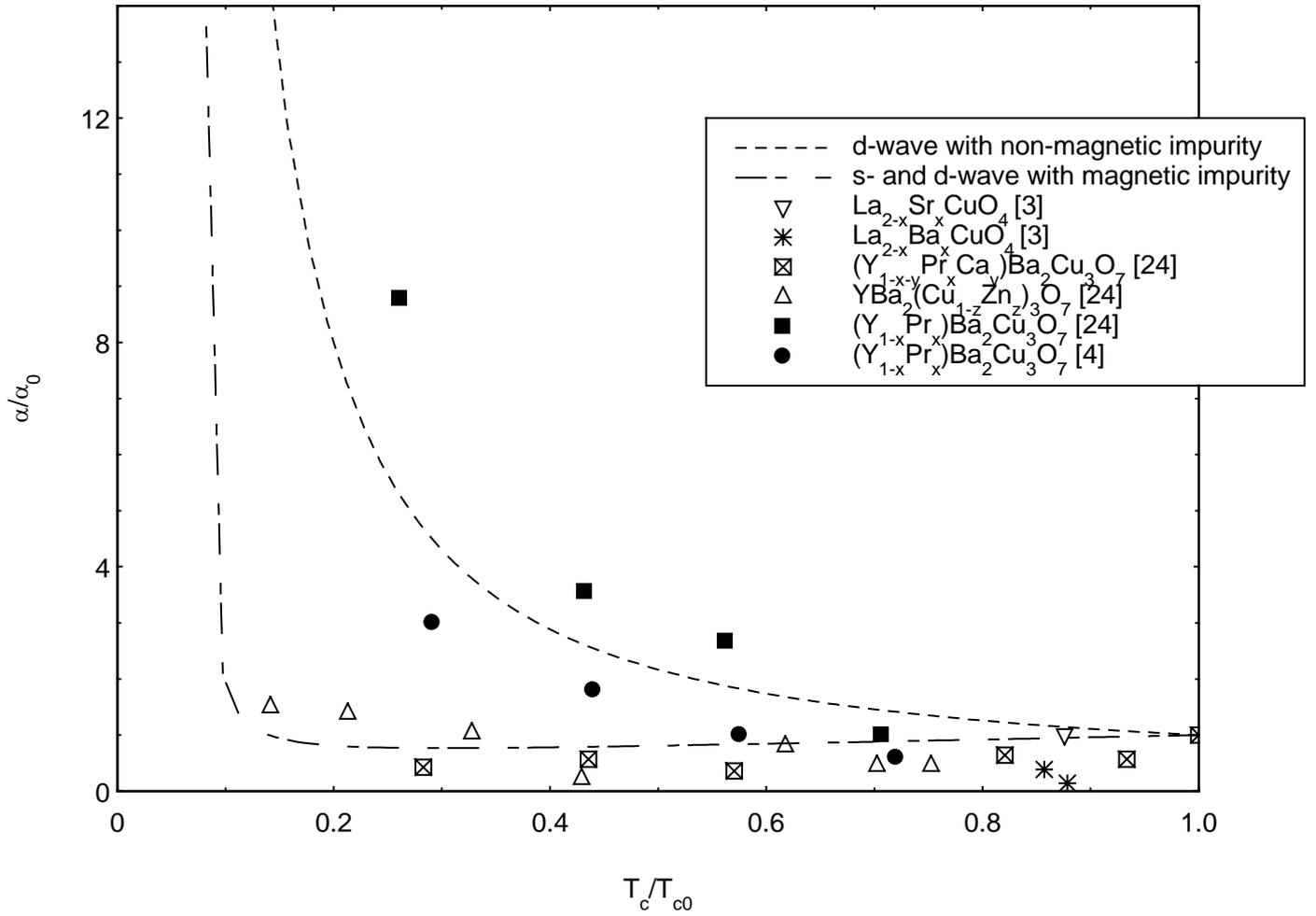

Figure. 1 .We show the numerical calculation of isotope coefficient fitted to experimental data [3,4,24].